\documentclass[twocolumn,showpacs,preprintnumbers,amsmath,amssymb]{revtex4}

\usepackage{graphicx}
\usepackage{dcolumn}
\usepackage{bm}

\begin{document}

\preprint{APS/123-QED}

\title{Universality in the one-dimensional chain of phase-coupled oscillators}

\author{Tony E. Lee}
\affiliation{Department of Physics, California Institute of Technology, Pasadena, CA 91125}
\author{G. Refael}
\affiliation{Department of Physics, California Institute of Technology, Pasadena, CA 91125}
\author{M. C. Cross}
\affiliation{Department of Physics, California Institute of Technology, Pasadena, CA 91125}
\author{Oleg Kogan}
\affiliation{Department of Physics and Astronomy, Michigan State University, East Lansing, MI 48824}
\author{Jeffrey L. Rogers}
\altaffiliation[Current address: ]{Microsystems Technology Office, Defense Advanced Research Projects Agency, Arlington, VA 22203}
\affiliation{Control and Dynamical Systems, California Institute of Technology, Pasadena, CA 91125}

\date{\today}

\begin{abstract}
We apply a recently developed renormalization group (RG) method to study synchronization in a one-dimensional chain of
phase-coupled oscillators in the regime of weak randomness. The RG predicts how
oscillators with randomly distributed frequencies and couplings form frequency-synchronized
clusters. Although the RG was originally intended for strong randomness,
i.e.\ for distributions with long tails, we find good agreement with numerical simulations
even in the regime of weak randomness. We use the RG flow to derive how the correlation length scales
with the width of the coupling distribution in the limit of large coupling. This leads to the identification of a universality class of 
distributions with the same critical exponent $\nu$.
We also find universal scaling for small coupling. Finally, we show that the RG flow is characterized by a universal approach to the unsynchronized fixed point, which provides physical insight into low-frequency clusters.
\end{abstract}

\pacs{}
\maketitle

\section{\label{sec:level1}Introduction}

There has been much interest in the spontaneous synchronization of nonlinear oscillators \cite{kuramoto84,pikovsky01,acebron05}.
Synchronization is the phenomenon of how oscillators with different intrinsic frequencies can oscillate at the same frequency
due to the interaction between them. As the coupling between them increases, the system becomes 
more synchronized, i.e. a given oscillator is more likely to be entrained with others. This phenomenon is found in many different contexts, including Josephson junctions \cite{wiesenfeld96}, lasers \cite{vladimirov03}, neural networks \cite{varela01}, chemical oscillators \cite{kuramoto84}, and even rhythmic applause \cite{neda00}.

Although systems that exhibit synchronization usually involve driven and dissipative dynamics, they may be analyzed using ideas from equilibrium statistical mechanics. For instance, a system can experience a phase transition from the unsynchronized to the synchronized state, in which a macroscopic fraction of the oscillators follows the same frequency. This is commonly known as the \emph{entrainment transition} \cite{hong05,hong07} and is analogous to the paramagnetic-ferromagnetic transition in spin models \cite{goldenfeld92}. The coupling between oscillators acts like the inverse temperature: as coupling increases, the system becomes more ordered.

In the mean-field case, when each oscillator is coupled to all others, the entrainment transition occurs at a finite value of the coupling \cite{kuramoto84,acebron05}. One may also consider finite-dimensional lattices of oscillators, and it has been shown that the lower critical dimension for entrainment is two \cite{sakaguchi87,daido88,hong07}. For $d\leq 2$, macroscopic entrainment exists only at infinite coupling, but for finite coupling, there are still local frequency-synchronized clusters \cite{strogatz88}. 

A previous work presented a real-space renormalization group (RG) approach for a one-dimensional chain of oscillators \cite{kogan09}.
It was successful at predicting cluster structure and frequency in the case of strong randomness, i.e.\ when
the frequency and coupling distributions had long tails. It was thought that strong randomness was
required in order for the perturbative decimation steps to be accurate. The numerical RG was successful over a wide range
of distribution widths and accurately predicted the scaling of the correlation length with distribution widths. An advantage of this RG approach was that it allowed the couplings to be randomly distributed. While most studies have assumed a constant coupling across the lattice, a physically realistic system would have disorder in the coupling \cite{acebron05}.

In this paper, we extend the analogy between synchronization and equilibrium physics even further by demonstrating universal features in the one-dimensional model. Universality is the existence of a class of systems that exhibit the same scaling behavior. For example, in spin models, the scaling of the correlation length with temperature near a critical point may be identical across different physical Hamiltonians. In general, the presence of universal behavior is important because it means that the physical properties being studied are in some ways independent of the microscopic details. In equilibrium physics, there is a close relationship between universality and the RG: the universal properties near a phase transition are explained by the RG flow there.

Here, we find a similar relationship between universality and the RG in the 1d nonequilibrium synchronization problem.
We first show that the RG is accurate over a wide range of frequency and coupling distributions, even in the regime
of weak randomness. Then by studying the flow of the RG for large and
small coupling, we find how the correlation length scales with the
width of the coupling distribution. This leads to the identification
of several distinct universality classes, based on generic features of the frequency and coupling distributions. In fact, universality exists
even far from the synchronized unstable fixed point. By studying the flow towards the unsynchronized stable fixed point, we find a universal approach to it, which dictates the dynamics of low-frequency clusters independently of
initial disorder realizations.

This paper is organized as follows. Section II reviews the RG and reports on its performance for weak randomness.
In Secs.~III and IV, we study the correlation length for large and small coupling, respectively. Section V examines the universal approach of the RG to the stable fixed point. In Sec.~VI, we conclude.

\section{\label{sec:level2}Renormalization group}
\subsection{Overview} \label{sec:rg_overview}
The one-dimensional chain of oscillators with nearest-neighbor interactions is described by the equations of motion:
\begin{eqnarray}
\dot{\theta_i}=\omega_i+K_{i-1}\sin(\theta_{i-1}-\theta_i)+K_i\sin(\theta_{i+1}-\theta_i) \;, \label{eq:eom}
\end{eqnarray}
where $\theta_i$ is the phase of the $i$th oscillator. The $\omega_i$ are the intrinsic frequencies
taken from a random distribution $\rho_{\omega}$, assumed to have zero mean without loss of generality. 
The $K_i$ are the couplings drawn from a random distribution $\rho_K$ and are assumed to be positive.
The couplings organize the oscillators into clusters of common frequency $\bar{\omega}$, defined as
\begin{eqnarray}
\bar{\omega}_i\equiv\lim_{(t-t_0)\rightarrow\infty}\frac{\theta_i(t)-\theta_i(t_0)}{t-t_0} \;. \label{eq:avg_freq}
\end{eqnarray}

In the renormalization group \cite{kogan09}, oscillators are combined into effective oscillators, so the model is slightly generalized to:
\begin{equation}
m_i\dot{\theta_i}=m_i\omega_i+K_{i-1}\sin(\theta_{i-1}-\theta_i)+K_i\sin(\theta_{i+1}-\theta_i) \;,
\end{equation}
where the parameter $m_i$ represents the number of original oscillators in the effective oscillator $i$. $m_i$ is referred
to as an oscillator's \emph{mass}, and the reduced mass between a pair of oscillators is given by 
$\mu=\frac{m_i m_{i+1}}{m_i+m_{i+1}}$.

The RG is based on two decimation steps, which correspond to a coarse-graining of the system. The first is the \emph{strong coupling decimation} step. Two oscillators, $n$ and $n+1$, connected by a large coupling $K_n$ would be expected to synchronize. Hence they are combined into a single effective oscillator
with mass $M$,
\begin{eqnarray}
M=m_n+m_{n+1} \;,
\end{eqnarray}
and intrinsic frequency $\Omega$,
\begin{eqnarray}
\Omega=(m_n\omega_n+m_{n+1}\omega_{n+1})/M \;.
\end{eqnarray}
This step is valid up to zeroth order in ratios such as $\omega_n/K_n$ and $K_{n-1}/K_n$. We refer to such a decimated pair of oscillators
as \emph{strongly coupled}. The new oscillator may continue to be recombined with other oscillators, and hence oscillators that are strongly coupled  belong to the same frequency-synchronized cluster. For the sake of clarity, when a pair of oscillators are decimated as strongly coupled, we say that their \emph{bond} has been decimated.

The second decimation step is the \emph{fast oscillator decimation} step. An oscillator with large intrinsic frequency $\omega_n$ relative to its neighbors is expected to rotate freely. Such an oscillator will not synchronize with its neighbors, so the coupling to the neighbors, $K_{n-1}$ and $K_n$, are set to zero. This step is valid up to first order in ratios such as $K_n/\omega_n$ and $K_{n-1}/\omega_n$. In practice, the fast oscillator is removed from the chain and stored for later analysis, and the coupling between neighbors $n-1$ and $n+1$ is set to zero. There is a second-order shift in intrinsic frequency for the fast oscillator and its neighbors \cite{kogan09}, but we ignore the shift in this paper because it is small. The fast oscillator may consist of multiple oscillators that were strongly coupled together. Hence, removal of a fast oscillator means that the cluster is in its final form.

The chain of oscillators is renormalized by successive application of the two decimation steps. The process is carried out numerically on a list of parameters $\{m_i,\omega_i,K_i\}$. To decide which oscillator or bond to decimate first, the energies of each are calculated: $\{|\omega_i|, \frac{K_i}{2\mu_{i,i+1}}\}$. \emph{Energy} in this context reflects how much influence something has on clustering. An oscillator with the highest energy in the chain has a large frequency and should probably be decimated as a fast oscillator. Similarly, a bond with the highest energy should probably be strongly coupled. The highest energies are decimated first, so that coarse-graining corresponds to decreasing the energy scale of the system. In practice, we decrement the energy scale $E$ and consider oscillators and bonds with energies $\geq E$ for decimation on each cycle.

Although an oscillator or bond may have the highest energy, it must satisfy another criterion involving its neighbors before being decimated. We calculate the ratio
\begin{eqnarray}
r_n\equiv\frac{K_n}{\mu_{n,n+1}|\omega_n-\omega_{n+1}|} \;,
\end{eqnarray}
which measures the tendency of a pair of oscillators to synchronize.
If $\frac{K_n}{2\mu} \geq E$ and $r_n>1$, the bond is strongly coupled. If $|\omega_n| \geq E$, $r_{n-1}<1$, and $r_n<1$, oscillator $n$ is removed as a fast oscillator. The threshold of 1 was chosen based on a numerical study of small chains \cite{kogan08} and is exact for a chain of two oscillators.

It is possible that an oscillator or bond is not immediately decimated despite having the largest energy in the chain. For instance, if $r_{n-1}<1$ but $r_n>1$, oscillator $n$ is not decimated as fast. Eventually though, the entire chain is decimated, leaving a list of fast oscillators. Each of these fast oscillators corresponds to a frequency-synchronized cluster. One may then study the statistics of cluster mass and frequency.

The clusters predicted by the RG are compared to those found by numerically integrating Eqs.~\eqref{eq:eom} with a variable stepping Runge-Kutta algorithm \cite{press92}.
To identify frequency clusters in the simulations, the average frequencies
are calculated according to Eq.~\eqref{eq:avg_freq}. A group of oscillators is determined to be a synchronized cluster if its members have the same value of $\bar{\omega}$ within some tolerance.

\subsection{\label{sec:level2b}Application to weak randomness}

In the previous work \cite{kogan09}, the RG was found to be in good agreement with
simulations in the regime of strong randomness.  Comparisons of
real-space cluster structure, cluster mass distribution, and cluster
frequency distribution were excellent. 
Distributions of both frequency ($\rho_{\omega}$) and coupling
($\rho_{K}$) were assumed to be Lorentzian.  The long tails of the Lorentzian distributions heightened the overall
randomness in the chain, helping to ensure accuracy of the
perturbative decimation steps and to support the intuitive notion of the
strong randomness case.  For instance, when decimating oscillator $n$ as
fast, $\omega_{n}$ should be much larger than the neighboring $\omega$,
$K$.

\begin{figure*}
\centering
\includegraphics[width=7 in]{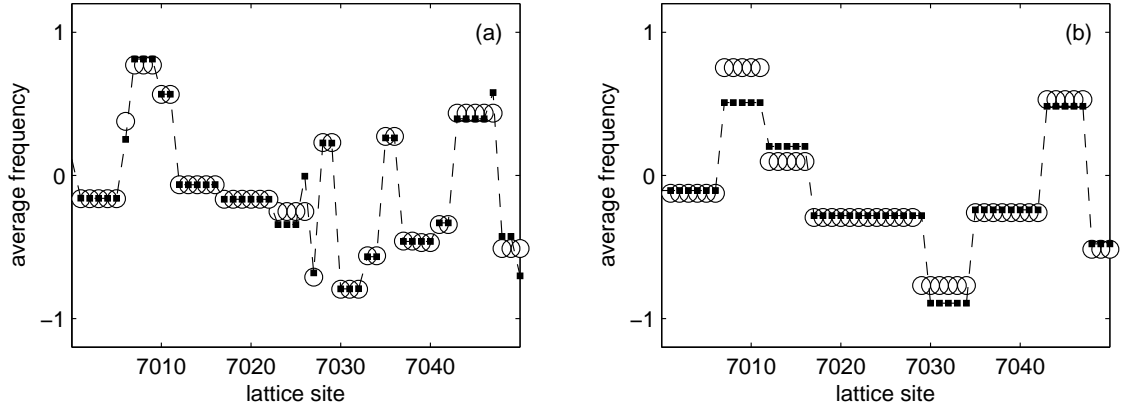}
\caption{\label{fig:realspace} Average frequency along the chain when $\omega$ and $K$ are both drawn from (a) rectangular and (b) triangular distributions. The RG predictions (open circles) are compared with simulation results (dashed lines, solid squares). The coupling width $\lambda$ is (a) 1 and (b) 10.}
\end{figure*}

\begin{figure*}
\centering
\includegraphics[width=7 in]{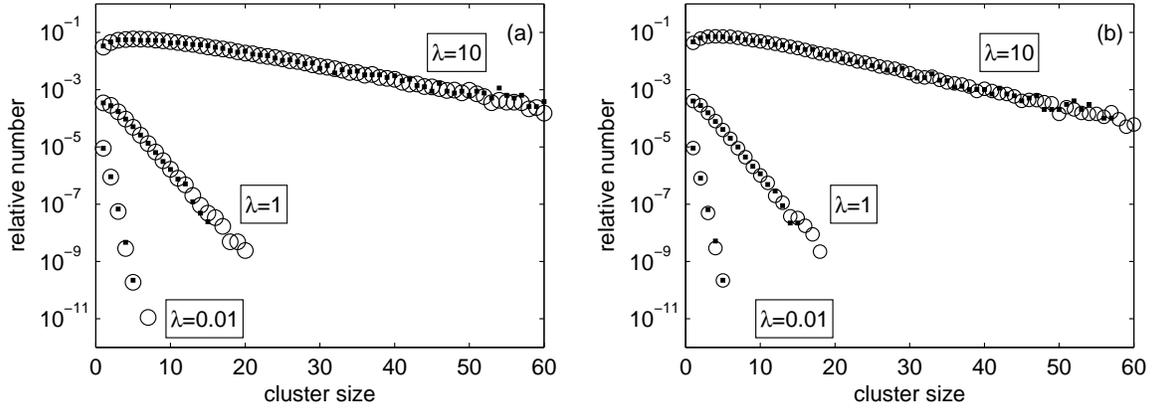}
\caption{\label{fig:histo_m} Cluster mass distribution when $\omega$ and $K$ are both drawn from (a) rectangular and (b) triangular distributions. The RG predictions (open circles) are compared with simulation results (solid squares). The plots for different values of the coupling width $\lambda$ are  offset for visibility.}
\end{figure*}

To understand the validity range of the RG, we relax the condition of strong
randomness by considering cases where $\rho_{\omega}$ and
$\rho_{K}$ have finite variance.  In particular, we use rectangular, triangular, and Gaussian distributions.
We still find good agreement with simulations in the regime of weak randomness. Figure \ref{fig:realspace} gives real-space comparisons between RG and simulation for rectangular and triangular distributions. Figure \ref{fig:histo_m} compares RG and simulation in terms of cluster mass distribution.  This shows that the RG is applicable over a wide range of distribution types and widths.

\begin{figure*}
\centering
\includegraphics[width=7 in]{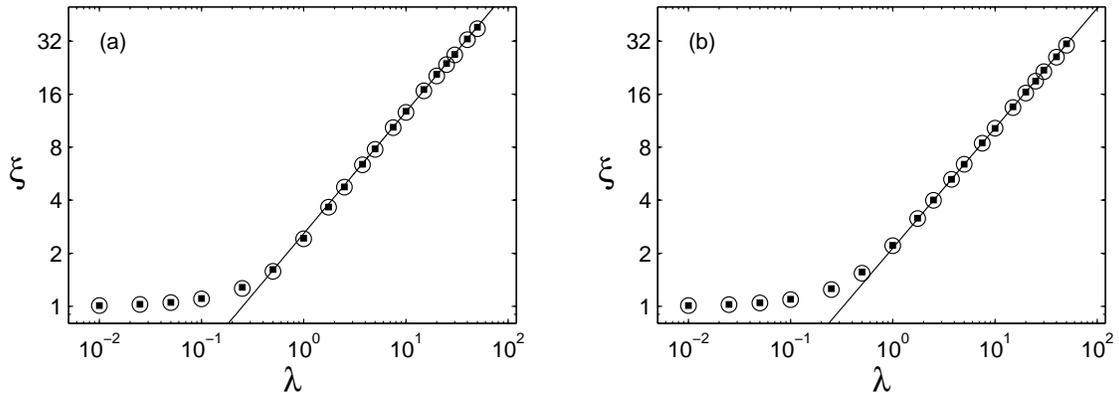}
\caption{\label{fig:xi_lambda} Correlation length $\xi$ vs.\ coupling width $\lambda$ when $\omega$ and $K$ are both drawn from (a) rectangular and (b) triangular distributions. The RG predictions (open circles) are compared with simulation results (solid squares). The lines show the power law fits to the solid squares with the exponent $\nu$ = 0.67. Here, $\xi$ is defined as the average cluster mass.}
\end{figure*}

The numerics were done with chains of $10^6$ and $10^5$ oscillators for RG and simulation, respectively.
$\rho_{\omega}$ is assumed to be symmetric around zero, while $\rho_{K}$ is the positive half of an otherwise symmetric distribution. The width of $\rho_{\omega}$ is defined to be the half width at half maximum for Lorentzian, triangular, and rectangular, and the standard deviation for Gaussian. The width of $\rho_{K}$ is defined similarly. The width of $\rho_{\omega}$ is set to one, without loss of generality, while the width $\lambda$ of $\rho_{K}$ is varied. The RG is in good agreement with simulations for a wide range of $\lambda$. In Sec.~\ref{sec:level5}, we provide insight into why the RG works so well even for weak randomness.

With the simulation data, one can study the shape of the mass distribution $\rho_m$. For large coupling width ($\lambda\geq 2.5$), it takes the form
\begin{eqnarray}
\rho_m(m) \sim m^{c_1} \,e^{-m/c_2} \;, \label{eq:rho_m_pow_exp}
\end{eqnarray}
where $c_1$ and $c_2$ are constants, and $c_1$ is approximately $2/3$ whenever $\rho_{\omega}$ and $\rho_K$ are of weak randomness. For smaller values of $\lambda$, the mass distribution has a similar but more complicated form.

In the limit of large coupling width, the correlation length $\xi$ of the system may be defined as the average cluster mass. In Fig.~\ref{fig:xi_lambda}, we plot $\xi$ as a function of $\lambda$ for rectangular and triangular distributions. For all distributions considered, the scaling
\begin{eqnarray}
\xi\sim\lambda^{\nu}
\label{eq:xivslambda}
\end{eqnarray}
holds for large $\lambda$, where $\nu$ is the critical exponent. Table \ref{tab:nu} lists the values of $\nu$ for different types of distributions. In Sec.~\ref{sec:level2c}, we comment on the significance of $\nu$. Note that $\lambda$ is a proxy for coupling strength, since larger $\lambda$ implies larger couplings in the system.

\begin{table}
\caption{\label{tab:nu}The critical exponent $\nu$, calculated by simulation, numerical RG, and analytical RG for different distribution types for $\omega$ and $K$. The first two columns list the distribution types: rectangular, triangular, Gaussian, Lorentzian, or delta function. The exponent $\nu$ describes how the correlation length scales with the coupling width: $\xi\sim\lambda^{\nu}$.}
\begin{ruledtabular}
\begin{tabular}{ccccc}
$\rho_{\omega}$ & $\rho_K$ & Simulation    & Numerical RG & Analytical RG \\
\hline
rec      & rec            & 0.671(8)       & 0.666(3)         & 2/3 \\
tri      & tri            & 0.673(8)       & 0.668(2)         & 2/3 \\
gau      & gau            & 0.69(1)        & 0.669(2)         & 2/3 \\
lor      & lor            & 0.47(1)       & 0.48(2)          & 1/2 \\
lor      & rec            & 0.51(2)       & 0.503(3)         & 1/2 \\
rec      & lor            & 0.668(8)      & 0.669(3)         & 2/3 \\
gau      & $\delta$       & 2.04(5)       & 2.003(4)          & 2 \\
\end{tabular}
\end{ruledtabular}
\end{table}

\subsection{\label{sec:level2c}Fixed points}

The RG described above is a \emph{functional RG} since the decimation steps modify distributions of $\omega$, $K$, and $m$, as opposed to a small set of parameters in a uniform system \cite{fisher94}. The flow of the RG is given by how those distributions evolve as energy decreases. This corresponds to a coarse graining due to the buildup and removal of clusters through the decimation steps. Since there is not global synchronization in one dimension \cite{strogatz88}, clusters cannot grow forever. Indeed, by the end of the RG, the chain has been completely divided into a list of fast oscillators. 

\begin{figure}
\centering
\includegraphics[width=3.5 in, viewport=25 655 300 730, clip]{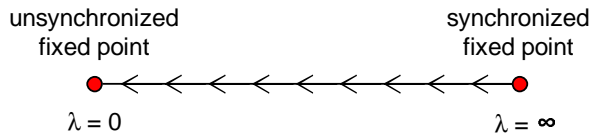}
\caption{\label{fig:flow_diagram} Flow diagram of the RG, showing the unsynchronized stable fixed point at $\lambda=0$ and synchronized unstable fixed point at $\lambda=\infty$.}
\end{figure}

The fixed points of the functional RG are the distributions of $\omega$ and $K$ that are invariant under the flow of the RG.
We identify two fixed points of the RG. The stable fixed point is given by $\lambda=0$, or $\rho_K(K)=\delta(K)$. This corresponds to the unsynchronized phase, since all the oscillators are freely rotating. The unstable fixed point is given by $\lambda=\infty$. This corresponds to the synchronized phase, since $\rho_{\omega}$ is then relatively narrowly peaked at zero. Any finite $\lambda$ will flow to the unsynchronized fixed point (Figs.~\ref{fig:flow_diagram} and \ref{fig:w_k2mu_flow_all}). See Sec.~\ref{sec:level5a} for further discussion of the unsynchronized fixed point.

The system may be interpreted as having a phase transition at $\lambda=\infty$, similar to the transition the one-dimensional Ising model has at zero temperature. The coupling width here is analogous to inverse temperature: the system becomes more ordered as $\lambda$ increases.
The scaling in Eq.~\eqref{eq:xivslambda} for large $\lambda$ probes the criticality of the system, and the correlation length $\xi$ diverges at the critical point. 

It is common for equilibrium systems to exhibit universality near the critical point. This means that macroscopic quantities, such as the correlation length, scale with temperature in the same way for different systems near the critical temperature \cite{goldenfeld92}. Although the model given by Eq.~\eqref{eq:eom} is far from equilibrium, since it is driven and overdamped, one may still find universal behavior near the critical point.
Indeed, Table \ref{tab:nu} shows that the exponent $\nu$ is the same across different types of disorder. When $\rho_{\omega}$ and $\rho_K$ are rectangular, triangular, or Gaussian, $\nu\approx 2/3$, and thus those systems belong to the same universality class. On the other hand, the Lorentzian case has $\nu\approx 1/2$ and is in a different universality class. 

The significance of universality is twofold. First, it means that the scaling given by $\nu$ does not depend on the microscopic details of a particular system. This is important when designing experiments to study synchronization. When it comes to observing the scaling in Eq.~\eqref{eq:xivslambda}, the exact shapes of $\rho_{\omega}$ and $\rho_K$ are not critical, as long as they are within a given universality class. Second, from a theoretical standpoint, universality is important in that it is often closely related to the RG flow near the critical point \cite{goldenfeld92}. In Sec.~\ref{sec:level3}, we draw the connection between universality and the RG flow. The fact that the RG correctly predicts the scaling laws boosts the claim that it is in fact a good representation of the model.

\section{\label{sec:level3}Universality for large coupling}

In this section, we provide an analytical understanding of the power-law scaling of the correlation length $\xi$ with the coupling width $\lambda$ in the regime of large $\lambda$. The exponent $\nu$ is seen to depend on generic features of the distributions of $\omega$ and $K$. Systems with the same such features will be in the same universality class. The approach here is similar to the usual one in equilibrium statisical mechanics, i.e.\ examining the RG flow near criticality.

\subsection{RG flow}

As explained in Sec.~\ref{sec:rg_overview}, oscillators and bonds are decimated in order of decreasing energy. To simplify the analysis in this section, we define the energy of a bond to be $K$ instead of $K/2\mu$. This means that a bond may be considered for strong coupling earlier on in the RG. This makes a difference only when the bond's $r\equiv K/\mu|\Delta\omega|>1$ when $E\equiv K$ but would have been $r<1$ when $E\equiv K/2\mu$. In practice, this happens rarely. Empirically, it is more accurate to use $K/2\mu$ as the energy, but using $K$ instead overestimates $\xi$ by only up to 5\%.

\begin{figure}
\centering
\includegraphics[width=3 in, viewport=40 600 260 750, clip]{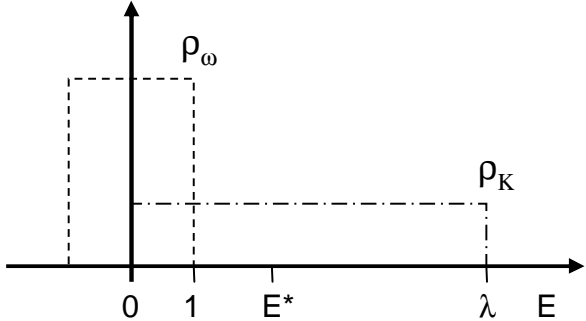}
\caption{\label{fig:cartoon_wk}An example of when the initial coupling distribution (dashed-dotted line) is wider than the initial frequency distribution (dashed line), i.e.\ when the coupling width $\lambda>1$. Rectangular distributions are shown here. The RG does mostly strong coupling decimation until some energy $E^*$, which corresponds to a length scale $\xi$ at which the system looks unsynchronized due to the emergence of fast oscillators.}
\end{figure}

When $\lambda\gg1$, most bonds have higher energy than all the oscillators (Fig.~\ref{fig:cartoon_wk}). Also, most bonds will satisfy $r>1$ and will be strongly coupled. Hence, the initial stage of the RG (when $E>E^*$ for some $E^*$) is given by the strong coupling of all bonds that satisfy $K\geq E$. Since the strong coupling decimation is done independently of the $m$ and $\omega$ of the corresponding oscillators, there are no correlations between oscillators remaining in the chain. $E^*$ is defined as the energy scale at which the presence of fast oscillators becomes important (when $r\simeq 1$). In the remainder of this section, we calculate the RG flow for $E>E^*$.

Let $\rho_m(m,E)$ and $\rho_K(K,E)$ be the normalized distributions of $m$ and $K$ of the chain at a given energy $E$. As the energy is decremented from $E$ to $E-dE$, \mbox{$dE\cdot\rho_K(E,E)$} pairs of oscillators are decimated as strongly coupled, and the corresponding couplings disappear. The flow of $\rho_K(K,E)$ is therefore given by simply rescaling on the interval $0<K<E$:
\begin{eqnarray}
\rho_K(K,E) = \frac{\rho_K(K,E_0)}{\int_0^{E} dK' \rho_K(K',E_0)} \;,
\end{eqnarray}
where $E_0$ is the initial energy and $\rho_K(K,E_0)$ is the initial $K$ distribution. The flow of $\rho_m$ is given by successively combining pairs of oscillators:
\begin{widetext}
\begin{eqnarray}
\rho_m(m,E-dE) = \frac{\rho_m(m,E) + dE\, \rho_K(E,E)\left[-2\rho_m(m,E)+\int\int dm_1 dm_2 \rho_m(m_1,E)\rho_m(m_2,E)\delta(m-(m_1+m_2))\right]}{1-dE \rho_K(E,E)} \;,
\end{eqnarray}
which leads to the integro-differential equation:
\begin{equation}
\frac{\partial\rho_m}{\partial E} = \rho_K(E,E) \left[\rho_m(m,E) - \int_0^m dm' \rho_m(m',E) \rho_m(m-m',E)\right] \;.
\end{equation}
\end{widetext}
We treat $m$ as a continuous variable on the range $[\,0,\infty)$. The initial condition at $E_0$ can be approximated by
\begin{eqnarray}
\rho_m(m,E_0)=e^{-m} \;,
\end{eqnarray}
which captures the fact that clusters have size one. The equation may be solved by Laplace transforming with respect to $m$. The solution is
\begin{eqnarray}
\rho_m(m,E) = \frac{1}{\ell(E)} e^{-m/\ell(E)}\;, \label{eq:rho_m}
\end{eqnarray}
where
\begin{eqnarray}
\ell(E) & = & e^{-\int_{E_0}^{E} dE' \rho_K(E',E')}\\
 & = & \frac{1}{\int_0^E dK' \rho_K(K',E_0)} \label{eq:etol}
\end{eqnarray}
is the average cluster mass at energy $E$. According to Eq.~\eqref{eq:rho_m}, $m$ is exponentially distributed in the initial stage of $E>E^*$. (The final distribution of cluster mass is not strictly exponential, as seen in Eq.~\eqref{eq:rho_m_pow_exp} and Fig.~\ref{fig:histo_m}.) Equation \eqref{eq:etol} relates energy to length scale. If $\rho_K(K,E_0)$ is continuous and nonzero near $K=0$,
\begin{eqnarray}
\ell \sim \frac{\lambda}{E} \qquad \mbox{when } E\ll\lambda \label{eq:LtoE}.
\end{eqnarray}
This is the case for triangular, Gaussian, rectangular, and Lorentzian distributions, although the proportionality constants differ.

\subsection{Correlation length}

To calculate the correlation length $\xi$, we find the energy scale $E^*$ at which the system looks unsynchronized. It is given by the point in the RG when $\langle r\rangle\simeq 1$, since that is when the RG starts to encounter a lot of fast oscillators, which end cluster formation. Hence,
\begin{eqnarray}
E^* \simeq \langle\mu|\Delta\omega|\rangle \;, \label{eq:exp_value}
\end{eqnarray}
where the expectation value is over all pairs of oscillators at energy $E^*$. Since there are no correlations between $m$ and $\omega$ of different oscillators, we can use Eq.~\eqref{eq:rho_m} for both oscillators in the pair. Also, an oscillator of mass $m$ has an $\omega$ given by the average of $m$ original frequencies. When $m\gg1$, we can apply the central limit theorem, to find the distribution
\begin{eqnarray}
\rho_{m,\omega}(m,\omega,E) & =&  \left(\frac{1}{\ell}e^{-m/\ell}\right) \left(\frac{1}{\sigma_0}\sqrt{\frac{m}{2\pi}}e^{-\frac{m\omega^2}{2\sigma_0^2}}\right) \; \label{eq:rho_mw_nonlor},
\end{eqnarray}
where $\sigma_0$ is the standard deviation of the original $\rho_{\omega}$. Thus the expectation value is given by
\begin{widetext}
\begin{eqnarray}
\langle\mu|\Delta\omega|\rangle & = & \int\int\int\int dm_1 dm_2 d\omega_1 d\omega_2 \,
\rho_{m,\omega}(m_1,\omega_1,E) \, \rho_{m,\omega}(m_2,\omega_2,E) \,\frac{m_1 m_2}{m_1 + m_2} \,|\omega_1 - \omega_2| \;.
\end{eqnarray}
\end{widetext}
The $\ell$ and $\sigma_0$-dependence can be removed from the integrand by scaling the integration variables appropriately, and the resulting integral can be done numerically:
\begin{eqnarray}
\langle\mu|\Delta\omega|\rangle & = & 0.42 \, \sigma_0 \, \ell^{1/2} \;. \label{eq:mudw_scaling}
\end{eqnarray}
The length scale at $E=E^*$ is given by $\ell(E^*)\equiv \xi$. Combining Eqs.~\eqref{eq:LtoE}, \eqref{eq:exp_value}, and \eqref{eq:mudw_scaling}, we find
\begin{eqnarray}
E^* & \sim & \xi^{1/2} \sim \left(\frac{\lambda}{E^*}\right) ^{1/2} \;. \label{eq:EtoXi}
\end{eqnarray}
Solving Eq.~\eqref{eq:EtoXi} for $E^*$ yields
\begin{eqnarray}
E^* & \sim & \lambda^{1/3} \label{eq:Estar}
\end{eqnarray}
and thereby
\begin{eqnarray}
\xi & \sim & \lambda^{2/3} \label{eq:xi_lambda}\;,
\end{eqnarray}
which predicts that $\nu=2/3$ and matches simulation results (Table \ref{tab:nu}). Although the exponent is the important quantity, we note that when all proportionality constants are included, the value of $\xi$ predicted here is within 10\% of the simulated result. We also note that $E^*$ is universal, according to Eq.~\eqref{eq:Estar}.

The universality of $\nu$ among Gaussian, triangular, and rectangular distributions is seen to come from Eqs.~\eqref{eq:LtoE} and \eqref{eq:rho_mw_nonlor}. In other words, if $\rho_{\omega}$ has finite variance and $\rho_K$ is continuous and nonzero at $K=0$, then $\nu=2/3$.

Now we check when the above argument is self-consistent. We assumed that $\lambda$ is large, so that the RG does only strong coupling decimation in the beginning. Also, $\xi$ should be large so that it is valid to treat $m$ as a continuous variable, to use the central limit theorem, and so that Eq.~\eqref{eq:LtoE} would hold at $E^*$. To find how large $\lambda$ should be for the power-law scaling to be accurate, we require $\xi\gg 1$ in Eq.~\eqref{eq:xi_lambda} in a self-consistent way. For rectangular $\rho_{\omega}$ and $\rho_K$, we find the condition $\lambda \gg 0.24$. According to simulation results in Fig.~\ref{fig:xi_lambda}, the power-law scaling holds for $\lambda\geq 7.5$.

One may use the above results to calculate the dynamical exponent $z$ for the system. It describes how the diverging time scale of the system scales with the diverging correlation length near criticality \cite{sondhi97}. Intuitively, when the chain is composed of long clusters, the clusters tend to have small frequencies, so the characteristic time scale of the system is large. The characteristic frequency of the final list of clusters may be approximated by the frequency standard deviation of the original oscillators at $E^*$:
\begin{eqnarray}
\langle\omega^2\rangle_o & = & \frac{\int\int dm \,d\omega \,\rho_{m,\omega}(m,\omega,E^*) \, m \, \omega^2}{\int dm \, \rho_m(m,E^*) \, m} \;.
\end{eqnarray}
The expectation value is done over the original oscillators, as opposed to clusters of oscillators, which is why the integrands include the factor $m$. In the case when Eq.~\eqref{eq:rho_mw_nonlor} holds, the time scale is given by
\begin{eqnarray}
\frac{1}{\sqrt{\langle\omega^2\rangle_o}} & \sim & \xi^{1/2} \;.
\end{eqnarray}
Thus, the dynamical exponent $z=1/2$ when $\rho_{\omega}$ has finite variance. This matches well with simulation results.

\subsection{Application to other distributions}

When $\rho_\omega$ is Lorentzian, $\sigma_0$ is infinite, so we cannot use the central limit theorem to characterize oscillator frequencies as in Eq.~\eqref{eq:rho_mw_nonlor}. Instead, we use the fact that the average of random variables drawn from a Lorentzian distribution is described by the same Lorentzian distribution. Thus, Eq.~\eqref{eq:rho_mw_nonlor} is modified to
\begin{eqnarray}
\rho_{m,\omega}(m,\omega,E) & =&  \left(\frac{1}{\ell}e^{-m/\ell}\right) \rho_\omega(\omega,E_0) \; \label{eq:rho_mw_lor},
\end{eqnarray}
so that the second factor is independent of $m$. Following the same procedure as before leads to the exponent \mbox{$\nu=1/2$}, which agrees well with simulations (Table \ref{tab:nu}). Note that \mbox{$\nu=2/3$} if $\rho_{\omega}$ is rectangular and $\rho_K$ is Lorentzian.

We can extend the analysis to the case where all the couplings are equal to a constant, here denoted by $\lambda$. This is the most commonly studied case in the literature. First we consider the situation when $\rho_K$ is a rectangle of width $\epsilon$ at $K=\lambda$:
\begin{eqnarray}
\rho_K(K,E_0) & = & \frac{1}{\epsilon} \qquad \lambda-\epsilon<K<\lambda \;.
\end{eqnarray}
Equation \eqref{eq:etol} gives
\begin{eqnarray}
\ell(E) & = & \frac{\epsilon}{E-\lambda+\epsilon} \qquad \lambda-\epsilon<E<\lambda \;.
\end{eqnarray}
If $\rho_{\omega}$ has finite variance, then Eq.~\eqref{eq:EtoXi} becomes
\begin{eqnarray}
E^* & \sim & \xi^{1/2} = \left(\frac{\epsilon}{E^*-\lambda+\epsilon}\right)^{1/2} \;,
\end{eqnarray}
which, in the limit of $\epsilon\rightarrow0$, leads to $\nu=2$, which is close to the simulated value 2.04(5). This agrees with the value given in \cite{hong07}, based on the linear approximation to Eq. \eqref{eq:eom}. The critical exponent can also be predicted using the result that for a chain of size $N$, the critical coupling for complete synchronization scales as $O(\sqrt{N})$ \cite{strogatz88,ochab09}. This implies that for an infinite chain, the length scale of synchronized clusters scales as $O(\lambda^2)$.

To summarize, we have derived the scaling behavior near the synchronized fixed point by studying the RG flow. The analytical predictions for the critical exponent match closely with the results of simulations and numerical RG (Table \ref{tab:nu}). We have presented a general procedure for handling different types of distributions, and it may be applied beyond the cases considered here. For example, one may study the case where $\rho_K$ diverges at $K=0$, which would require the scaling in Eq.~\eqref{eq:LtoE} to be modified.

\section{\label{sec:level4}Universality for small coupling}

In this section, we calculate the correlation length $\xi$ for small coupling and show that universality also exists in this regime. This is surprising, since universality is usually found in the vicinity of a critical point. This discussion generalizes the estimate for $\xi$ given in Ref.~\onlinecite{kogan09} to non-Lorentzian distributions and is cast from an RG point of view.

Since the average cluster mass approaches one in the limit of $\lambda\rightarrow 0$, we use a different definition for the correlation length based on the exponential decay of the final mass distribution $\rho_m(m)$:
\begin{eqnarray}
e^{-1/\xi} = \frac{\rho_m(2)}{\rho_m(1)}\label{eq:xi_def}
\end{eqnarray}
It is not clear if $\rho_m$ has the same exponential form for all $m$, because it is difficult to collect simulation statistics for clusters made of more than two oscillators for $\lambda\ll 1$. In this regime, there are very few such clusters. Hence, the above definition is restricted to $m=1,2$. Now, $\xi$ can be less than one and in fact approaches zero when $\lambda\rightarrow 0$.

The RG flow can be used to calculate $\xi(\lambda)$ as in the previous section. 
Since $\rho_K$ is now much narrower than $\rho_{\omega}$ at the beginning of the RG, most of the oscillators will be decimated as fast oscillators. Only when $E\leq\lambda$ will bonds start to be strongly coupled. If a bond satisfies $r\equiv K/\mu|\Delta\omega|>1$ initially, it will still be so when the energy is at the bond's energy, because removing neighboring fast oscillators does not affect this bond's $r$. Since a cluster of mass $m$ requires a consecutive sequence of \mbox{$m-1$} strongly coupled bonds, $e^{-1/\xi}$ in Eq.~\eqref{eq:xi_def} is equal to the probability of $r>1$ given by the initial frequency and coupling distributions:
\begin{eqnarray}
e^{-1/\xi} & = & \int_0^{\infty} dK\, \rho_K(K) \int_0^{2K} d\Delta\, \rho_{\Delta}(\Delta) \label{eq:alpha}\;,
\end{eqnarray}
where we have set $\mu=1/2$, and
\begin{eqnarray}
\hspace{-5mm}\rho_{\Delta}(\Delta) & = & 2\int d\omega\, \rho_{\omega}(\omega)\,\rho_{\omega}(\omega-\Delta) \quad\quad \Delta\geq 0 \;
\end{eqnarray}
is the distribution of the absolute value of frequency differences between neighboring oscillators.

If $\rho_K$ does not have long tails, the integral over $\Delta$ is approximately $2K\rho_{\Delta}(0)$. This gives
\begin{eqnarray}
e^{-1/\xi} & = & 4\,\langle K\rangle\int_{-\infty}^{\infty} d\omega\, \rho_{\omega}(\omega)^2 \label{eq:alpha_approx}\\
 & = & c\,\lambda \;,
\end{eqnarray}
where the proportionality constant $c$ depends on the distribution types.
The correlation length is then:
\begin{eqnarray}
\xi & = & -\frac{1}{\log\lambda + \log c} \\
 & \approx & -\frac{1}{\log\lambda}\;. \label{eq:xi_smalllambda}
\end{eqnarray}
For very small $\lambda$, the constant $c$ drops out, and the form of $\xi(\lambda)$ is universal across Gaussian, triangular, and rectangular distributions.

On the other hand, if $\rho_K$ has long tails such that $\langle K\rangle$ is infinite, the approximation in Eq.~\eqref{eq:alpha_approx} does not hold, and Eq.~\eqref{eq:alpha} must be integrated directly. For example, when $\rho_{\omega}$ and $\rho_K$ are both Lorentzian \cite{kogan09},
\begin{eqnarray}
e^{-1/\xi} & = & \frac{4\lambda}{\pi^2}\int_0^{\infty}dK \; \frac{\;\arctan K\;}{\lambda^2+K^2} \; \label{eq:alpha_lor},
\end{eqnarray}
and the integral may be done numerically.

\begin{figure}
\centering
\includegraphics[width=3.5 in]{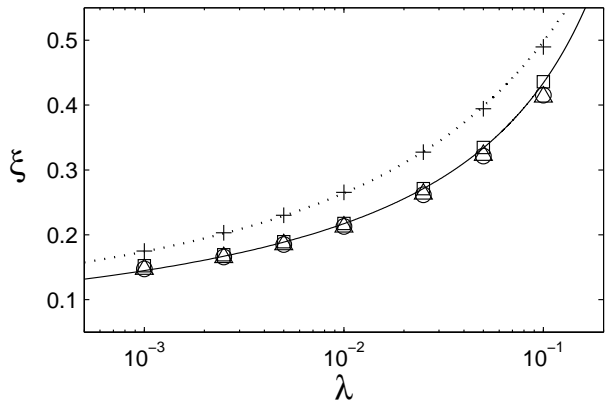}
\caption{\label{fig:xi_smalllambda} Correlation length $\xi$ vs.\ coupling width $\lambda$ for small $\lambda$ for Gaussian (circles), triangular (triangles), rectangular (squares), and Lorentzian (plus signs) distributions, calculated from simulations. The prediction from analytical RG is plotted for weak randomness (solid line) and Lorentzian (dotted line). Lorentzian is in a different universality class from the others. Here, $\xi$ is defined according to Eq.~\eqref{eq:xi_def}.}
\end{figure}

Figure \ref{fig:xi_smalllambda} shows that the predictions in Eqs.~\eqref{eq:xi_smalllambda} and \eqref{eq:alpha_lor} agree well with simulations for the various distribution types. It is clear that Lorentzian is in a different universality class for small $\lambda$. Note that the universality class is determined by whether $\langle K\rangle$ is finite: if $\rho_{\omega}$ is Lorentzian while $\rho_K$ is Gaussian, the scaling in Eq.~\eqref{eq:xi_smalllambda} still holds.

\section{\label{sec:level5}Universal Approach to the Stable Fixed Point}

In this section, we examine the flow of the RG near the unsynchronized fixed point. The discussion here is different from that given in Sec.~\ref{sec:level4}, where we assumed that the system started out already near the unsynchronized fixed point, i.e.\ coupling width $\lambda\ll 1$. Here, we allow the system to start anywhere, including $\lambda \gg 1$, and look at the flow of the RG after enough renormalization steps have been carried out so that the system approaches the unsynchronized fixed point.

We return to using $K/2\mu\equiv\tilde{K}$ as the bond energy instead of just $K$.
 
\subsection{\label{sec:level5a}Unsynchronized stable fixed point}

\begin{figure}
\centering
\includegraphics[width=3 in, viewport= 30 620 310 740,clip]{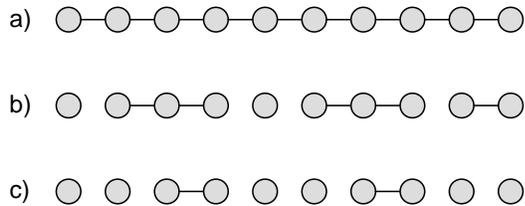}
\caption{\label{fig:chain_evo} As the RG progresses, the chain is punctured by more and more $K=0$ bonds. At the start of the RG (a), all oscillators are coupled to their neighbors. Near the end of the RG (c), it is unlikely to have two nonzero bonds in a row. We study the properties of pairs of oscillators that are still connected with a nonzero bond.}
\end{figure}

\begin{figure*}
\centering
\includegraphics[width=7 in]{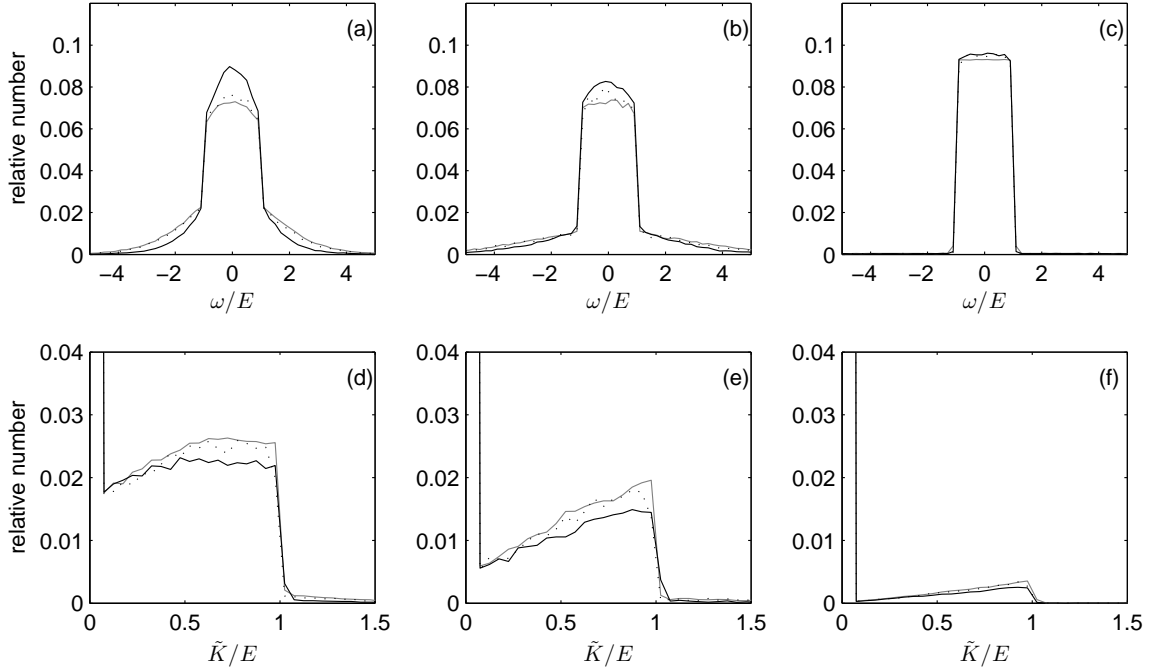}
\caption{\label{fig:w_k2mu_flow_all}Distribution of oscillator frequency (top) and bond energy (bottom) at different stages of the RG: (a,d) 50\%, (b,e) 25\%, and (c,f) 3\% of original oscillators left in chain. The results for different initial distributions of $\omega$ and $K$ are shown: Lorentzian $\lambda=7.5$ (black, solid), triangular $\lambda=1.25$ (gray, solid), triangular $\lambda=7.5$ (black, dotted). The RG approaches the fixed point regardless of the initial chain: the distribution of $\omega$ becomes rectangular, while the distribution of $\tilde{K}$ approaches a delta function at zero.}
\end{figure*}

At the unsynchronized fixed point, $\rho_K(K)=\delta(K)$, meaning that all the remaining oscillators are effectively uncoupled to each other. Since this fixed point is stable, at late stages of the RG, the chain will be punctured by many $K=0$ bonds (Fig.~\ref{fig:chain_evo}). This is due to the removal of a lot of fast oscillators, which leaves the former neighbors uncoupled. Due to the large fraction of $K=0$ bonds, most oscillators will be uncoupled to both neighbors. Such isolated oscillators are waiting to be removed as fast oscillators and will not be strongly coupled with their neighbors. Unless the $\omega$ distribution of isolated oscillators diverges at zero, it will approach a uniform distribution on the interval $[-E,E]$ and zero elsewhere, since the flow is given by successively chopping off the sides. Thus at the unsynchronized fixed point, $\rho_{\omega}(\omega)=\frac{1}{2E}$ for $|\omega|\leq E$. Note that these isolated, low-frequency oscillators can be clusters of mass one or of higher mass, depending on the size of $\lambda$.

\begin{figure*}
\centering
\includegraphics[width=7 in]{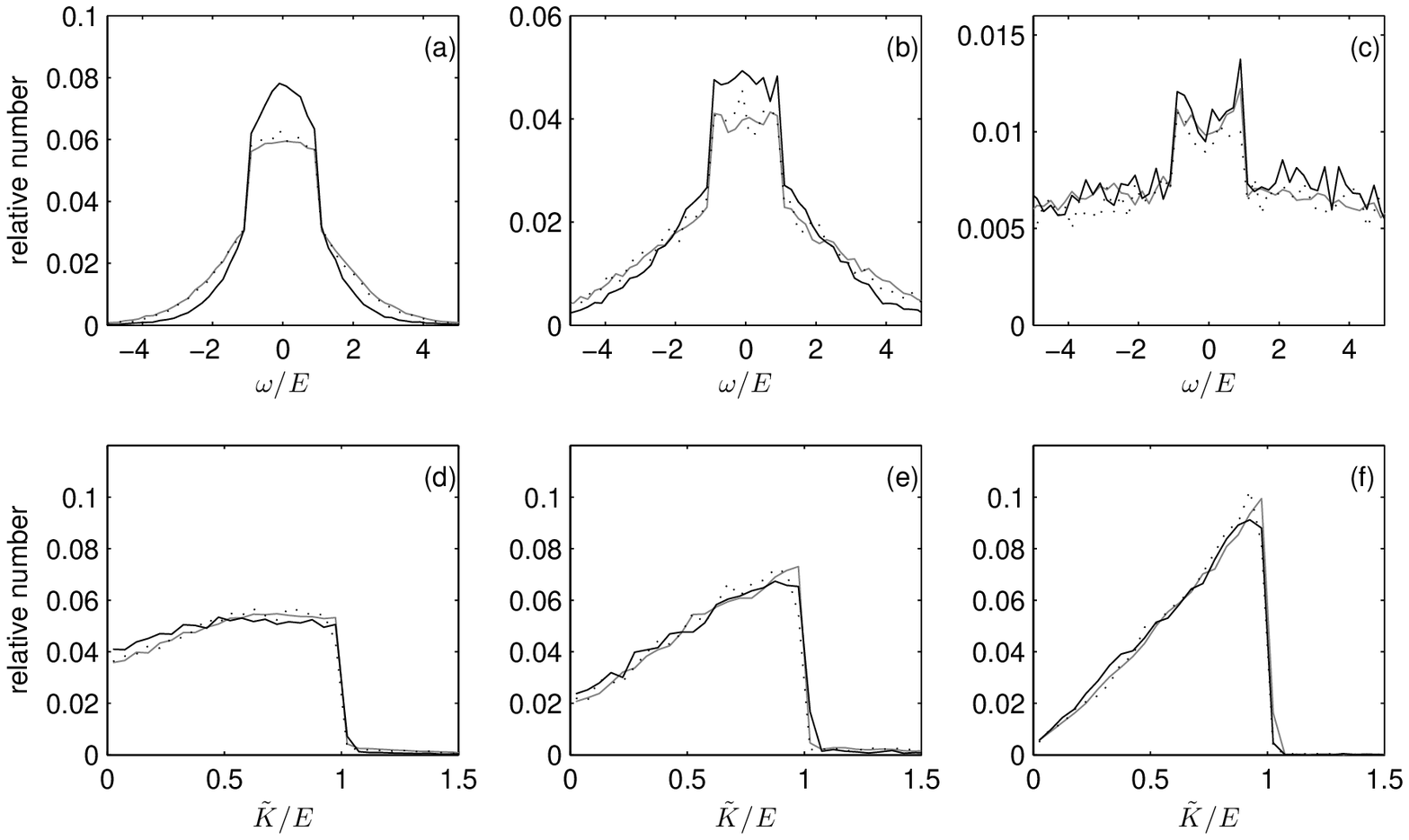}
\caption{\label{fig:w_k2mu_flow_iso}Distribution of frequency of unisolated oscillators (top) and nonzero bond energies (bottom) at different stages of the RG: (a,d) 50\%, (b,e) 25\%, and (c,f) 3\% of original oscillators left in chain. The results for different initial distributions of $\omega$ and $K$ are shown: Lorentzian $\lambda=7.5$ (black, solid), triangular $\lambda=1.25$ (gray, solid), triangular $\lambda=7.5$ (black, dotted). These are the perturbations that decay as the RG approaches the fixed point. The $\omega$ perturbation becomes quadratic in the middle and flat otherwise, while the $\tilde{K}$ perturbation becomes triangular. This shows that the system approaches the fixed point in the same way, regardless of the initial chain.}
\end{figure*}

Due to the stability of the unsynchronized fixed point, almost all chains will flow to it, regardless of the initial distributions of $\omega$ and $K$. Figures \ref{fig:w_k2mu_flow_all} shows examples of chains with different initial distributions flowing to the same fixed point.

Now we consider the way $\rho_{\tilde{K}}$ and $\rho_{\omega}$ approach their respective fixed distributions. We look at the component of the distributions that are not the fixed distributions. For $\rho_{\tilde{K}}$, this means considering only bonds with $\tilde{K}\neq 0$. For $\rho_{\omega}$, this means considering only oscillators with at least one nonzero bond. These components are perturbations that decay as the RG approaches the fixed point. Figures \ref{fig:w_k2mu_flow_iso} show these perturbations.

It is seen that the $\tilde{K}$ perturbation collapses onto a triangular distribution regardless of the initial distributions, while the $\omega$ perturbation collapses onto a flat and quadratic shape. Hence, there is a universal approach to the stable fixed point. One can think of the fixed point as having a least negative eigenvalue, corresponding to a favored direction for approaching the fixed point in the functional RG space.

This suggests why the RG works well even with weak randomness. Regardless of whether the chain starts with strong or weak randomness, distributions of oscillator properties will look similar as the RG progresses. 

In the following section, we explain how the triangular $\tilde{K}$ perturbation comes about. Due to rather technical details, we postpone the explanation of the flat-quadratic $\omega$ perturbation to the appendix.

\subsection{Explanation of the triangular perturbation} \label{sec:tri_pertubation}

Here, we explain how the RG algorithm leads to the universal shape of the $\tilde{K}$ perturbation. Since the perturbation is due to nonzero bonds, we consider pairs of oscillators with a nonzero bond connecting them, but otherwise uncoupled to the rest of the chain (Fig.~\ref{fig:chain_evo}). We ignore single oscillators that are uncoupled to both neighbors, since they are waiting to be removed as fast oscillators and do not contribute to the buildup of future clusters.

To facilitate the discussion, we define a \emph{center} to be a pair of oscillators that both satisfy $|\omega|<E$, where $E$ is the energy of the RG at a given stage. A \emph{tail} is a pair of oscillators where at least one $|\omega|>E$. In other words, neither of the oscillators in a center has been checked for fast oscillator decimation. On the other hand, at least one of the oscillators in a tail has been checked for fast oscillator decimation, but since it is still in the chain, we know that it satisfies $r>1$ and is waiting to be strongly coupled (when $E=\tilde{K}$).

A pair can stop being a center by either being checked for fast oscillator decimation or strong coupling decimation. In the first case, the pair will either be removed ($r<1$) or become a tail ($r>1$). In the second case, the pair will either be strongly coupled ($r>1$) or remain intact in the chain ($r<1$). It can be shown that when a center gets checked for strong coupling decimation, it will always be decimated \cite{proof1}. It is also possible for new centers to be formed, due to the strong coupling of other oscillators; however, we can neglect this possibility when it is unlikely to have two nonzero bonds in a row.

Let the centers be described by the distribution $\rho^c(\omega_1,\omega_2,\tilde{K};E)$, where $|\omega_1|>|\omega_2|$. This distribution is nonzero only when $|\omega_1|,\tilde{K}<E$. Its flow is given by setting to zero the region that doesn't satify that inequality, and then normalizing to unity. Unless there is a divergence at the origin, the distribution of centers approaches
\begin{equation}
\rho^c(\omega_1,\omega_2,\tilde{K};E)=\frac{1}{2E^3} \quad\quad |\omega_2|<|\omega_1|<E, 0<\tilde{K}<E \;. \label{eq:rho_c_full}
\end{equation}
This is just the statement that any smooth distribution will look flat if you keep chopping off its sides. Integrating out $\omega_1,\omega_2$ in Eq.~\eqref{eq:rho_c_full}, we find that the distribution of $\tilde{K}$ of centers is flat:
\begin{eqnarray}
\rho^c_{\tilde{K}}(\tilde{K};E) & = & \frac{1}{E} \quad\quad 0<\tilde{K}<E \;. \label{eq:rho_c_k2mu}
\end{eqnarray}

\begin{figure}
\centering
\includegraphics[width=3 in, viewport= 30 580 290 750,clip]{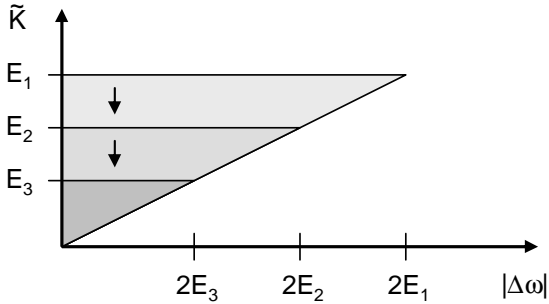}
\caption{\label{fig:rho_tail_flow} Illustration of the flow of tail pairs according to Eq.~\eqref{eq:tails}. The distribution is constant on a triangular region in $|\Delta\omega|,\tilde{K}$ space and zero elsewhere. As energy decreases, the triangular region shrinks, but keeps the same shape. Both existing tails and incoming tails at a given energy are uniformly distributed on the same triangular region.}
\end{figure}

We now consider centers at the moment they get checked for fast oscillator decimation, i.e.\ when $\omega_1=E$. Here we assume that $\omega_1$ is positive, but analogous results hold for when it is negative.
\begin{equation}
\rho^c_{\omega_1=E}(\omega_2,\tilde{K};E)=\frac{1}{2E^2} \quad\quad 0<|\omega_2|,\tilde{K}<E \;
\end{equation}
At this point, if $\omega_2$ and $\tilde{K}$ are such that $r<1$, then oscillator 1 will get removed as a fast oscillator, and oscillator 2 will become an isolated oscillator, which is then ignored. But if $r>1$, then the pair becomes a tail and remains in the chain. Thus the new tails at energy E are described by:
\begin{equation}
\rho^t_{\omega_1=E}(\omega_2,\tilde{K};E)=\frac{1}{E^2} \quad\quad 0<\frac{|\omega_2-E|}{2}<\tilde{K}<E \;,
\end{equation}
which can be rewritten as
\begin{equation}
\rho^t_{\omega_1=E}(|\Delta\omega|,\tilde{K};E)=\frac{1}{E^2} \quad\quad 0<\frac{|\Delta\omega|}{2}<\tilde{K}<E \;. \label{eq:tails}
\end{equation}
This distribution is constant over a triangular region in $|\Delta\omega|$, $\tilde{K}$ as illustrated in Fig.~\ref{fig:rho_tail_flow}.

Now here is the key. Tail pairs remain in the chain until they are decimated as strongly coupled, which happens when the RG energy reaches the bond energy. Thus Eq.~\eqref{eq:tails} describes the flow of the existing tails at any energy $E$. Existing tails have the same distribution as incoming tails at any energy, so Eq.~\eqref{eq:tails} describes all tails.

By integrating out $|\Delta\omega|$ in Eq.~\eqref{eq:tails}, one finds that the $\tilde{K}$ distribution of tails is triangular:
\begin{eqnarray}
\rho^t_{\tilde{K}}(\tilde{K};E) & = & \frac{2}{E^2}\;\tilde{K} \quad\quad 0<\tilde{K}<E \;. \label{eq:rho_t_k2mu}
\end{eqnarray}
In principle, the combined $\tilde{K}$ distribution of center and tail pairs is given by a weighted average of Eqs.~\eqref{eq:rho_c_k2mu} and \eqref{eq:rho_t_k2mu}. But it is shown in the appendix that as energy decreases, there are more and more tails relative to centers, so that at low energy, all non-zero $\tilde{K}$ are given by just Eq.~\eqref{eq:rho_t_k2mu}. This explains the triangular distribution in Fig.~\ref{fig:w_k2mu_flow_iso}.

The universality of Eq.~\eqref{eq:rho_t_k2mu} comes from the applicability of Eq.~\eqref{eq:rho_c_full}. Unless $\rho_{\omega}$ diverges at $\omega=0$, $\rho_{\tilde{K}}$ will approach $\delta(\tilde{K})$ through a decaying triangular perturbation as energy decreases. This explains the universal approach to the stable fixed point.

\subsection{Physical interpretation}

The above discussion has focused on RG flow at low energies, which corresponds to low frequency clusters. When $E$ is sufficiently small, Eq.~\eqref{eq:rho_c_full} accurately describes centers. In physical terms, Eq.~\eqref{eq:rho_c_full} describes pairs of clusters with $|\omega_1|,|\omega_2|<E$ that are interacting via a bond $\tilde{K}$. Depending on how big the coupling is, the two clusters may end up synchronizing and forming a bigger cluster.

Most of the clusters with $|\omega|<E$ will be effectively isolated from the rest of the chain, since most couplings are zero at low energy. It is shown in the appendix that of the cluster pairs that \emph{are} coupled and have at least one cluster with $|\omega|<E$, 3/5 are centers and 2/5 are tails. The centers have interactions given by Eq.~\eqref{eq:rho_c_k2mu}, while tails have interactions given by \eqref{eq:rho_t_k2mu}. Since Eqs.~\eqref{eq:rho_c_full} and \eqref{eq:rho_t_k2mu} and the proportion of centers and tails are universal across different initial distributions, we know that the dynamics at low frequency are also universal. Thus, this analysis of the RG provides insight into the interactions of low frequency clusters.

A possible extension of this analysis is to study the distribution of cluster mass. As mentioned in Sec.~\ref{sec:level2b}, the final distribution of cluster mass is of the form in Eq.~\eqref{eq:rho_m_pow_exp}. In the current discussion, the mass information is stored in the bond energy $K/2\mu$. We have considered the flow of pairs of oscillators that are uncoupled from the rest of the chain. By considering the flow of longer isolated chains, one may be able to understand the buildup of larger clusters and hence the final mass distribution. The difficulty lies in the correlations that appear due to the $r$ criterion.

\section{\label{sec:level6}Conclusion}

In this paper, we have explored various features of the real-space RG
approach to 1d synchronization, first presented in
Ref.~\onlinecite{kogan09}. We have shown that the RG method performs
well even beyond the strong randomness case, for which it was
originally intended. The RG was also used to calculate critical
properties of random oscillator chains, such as 
correlation-length scaling for both large and small coupling. We
identified several universality classes, whose behavior we also derived
analytically. Excellent agreement was found between our
analytical arguments, numerical RG, and simulations. Finally, we
demonstrated the universal approach to the stable fixed point.

The results presented here may find relevance in physical realizations
of the one-dimensional model. The universality implies that the
predicted scaling could be exhibited in experimental realizations of
the model without fine tuning: as long as the distributions of $\omega$ and $K$
have the generic features of a given universality class, the
corresponding scaling laws will hold. In addition, the dynamical
exponent derived here shows the relationship between time and length
scales in this model.

We emphasize that the results were all based on the RG as opposed to a traditional dynamical-systems approach. A natural next step is to develop an RG for higher dimensional lattices. It has been determined numerically that the lower critical dimension for macroscopic entrainment is two \cite{hong05,hong07}. Thus for $d>2$, an RG would have two stable fixed points, corresponding to the synchronized and unsynchronized phases \cite{goldenfeld92}. Once an RG has been developed for higher dimension, it may be possible to obtain an analytical understanding of synchronization in a way similar to the present work. By studying the RG flow near criticality, one may even find universal behavior in higher dimension.

This work has been supported by Boeing. GR thanks the Research
Corporation and the Packard foundation for their generous support. 
We also thank Heywood Tam for many helpful discussions.

\appendix

\section{Flow of tails and centers}
In this appendix, we calculate in more detail the flow of distributions of center and tail pairs. We consider late stages of the RG when it is unlikely to have two nonzero couplings in a row. We look at pairs of oscillators with a nonzero coupling between them, but otherwise uncoupled to the rest of the chain. Let $\omega_1$ and $\omega_2$ be the frequencies of the pair, with $|\omega_1| > |\omega_2|$. A center pair satisfies $|\omega_1|,|\omega_2|<E$, while a tail pair satisfies $|\omega_1|>E$. Thus, at least one of the oscillators within a tail pair has been checked for fast oscillator decimation and failed. Since we are considering only center and tail pairs, once such a pair gets decimated as either strongly coupled or fast, it drops out of the discussion.

We consider the rates of three processes as energy is decreased: the rate at which centers are decimated as strongly coupled, the conversion rate of centers into tails, and the rate at which tails are decimated as strongly coupled. In general, the distributions will not be normalized to one, since we would like to track the number density as energy is decreased.

We repeat part of the discussion in Sec. \ref{sec:tri_pertubation} with generalized notation for the sake of clarity. The starting energy is denoted by $\tilde{E}$, while the energy at a given stage of the RG is $E$. The energy at which a particular tail pair is created is denoted by $E_0$. Note that energies are always taken to be positive. 

\subsection{Flow of centers}

We start the calculation by assuming that the distribution of all unisolated pairs is given by the fixed point distribution of centers, Eq.~\eqref{eq:rho_c_full}.  This means that all pairs are centers, and there are no tails. The number density of tails is
\begin{equation}
\rho^c(\omega_1,\omega_2,\tilde{K};E)=\frac{1}{2\tilde{E}^3} \quad\quad |\omega_2|<|\omega_1|<E, 0<\tilde{K}<E \;. \label{eq:rho_c_full_appendix}
\end{equation}
This distribution is normalized to $(E/\tilde{E})^3$. The number of centers decreases as $E$ decreases due to the decimation of centers as strongly coupled or fast and the conversion of centers into tails. Note that we assume that no new centers are formed, which is a valid assumption during late stages of the RG when most couplings are zero.

By integrating out variables in Eq.~\eqref{eq:rho_c_full_appendix}, one obtains the distributions of $\omega_1$ and $\omega_2$:
\begin{eqnarray}
\rho^c_{\omega_1}(\omega_1;E) & = & \frac{E}{\tilde{E}^3} |\omega_1| \quad\quad\quad\quad\quad |\omega_1|<E  \label{eq:rho_c_w1} \\
\rho^c_{\omega_2}(\omega_2;E) & = & \frac{E}{\tilde{E}^3} (E-|\omega_2|) \quad\quad |\omega_2|<E \;. \label{eq:rho_c_w2}
\end{eqnarray}
Thus the number density of all center frequencies is
\begin{eqnarray}
\rho^c_{\omega}(\omega;E) & = & \rho^c_{\omega_1}(\omega;E) + \rho^c_{\omega_2}(\omega;E) \\
 & = & \frac{E^2}{\tilde{E}^3} \quad\quad |\omega|<E \;, \label{eq:rho_c_w}
\end{eqnarray}
which is normalized to $2(E/\tilde{E})^3$, reflecting the fact that each center contributes two frequencies.

\subsection{Flow of tails}

Now we consider the transition from center to tail, i.e.\ when $\omega_1=E$. We assume for now that $\omega_1$ is positive, but analogous results hold for when it is negative. Immediately before checking the pair for fast oscillator decimation, the pair is described by

\begin{eqnarray}
\rho^c_{\omega_1=E_0}(\omega_2,\tilde{K};E_0) & = & \frac{\rho^c(E_0,\omega_2,\tilde{K};E_0)}{\rho^c_{\omega_1}(E_0;E_0)} \\
 & = & \frac{1}{2E_0^2} \quad\quad 0<|\omega_2|,\tilde{K}<E_0 \;.
\end{eqnarray}
There is a half chance that the pair will satisfy $r<1$ and be decimated out, and there is a half chance that it will satisfy $r>1$ and become a tail. Assuming the latter case, immediately after becoming a tail, the pair is described by
\begin{equation}
\rho^t_{\omega_1=E_0}(\omega_2,\tilde{K};E_0)=\frac{1}{E_0^2} \quad\quad 0<\frac{|\omega_2-E_0|}{2}<\tilde{K}<E_0 \;,
\end{equation}
which reflects the fact that $r>1$. As shown in Fig.~\ref{fig:rho_tail_flow}, the flow of tails created at $E_0$ as energy $E$ decreases is given by
\begin{equation}
\rho^t(\omega_2,\tilde{K};E,E_0)=\frac{1}{E_0^2} \quad\quad 0<\frac{|\omega_2-E_0|}{2}<\tilde{K}<E \;.
\end{equation}

By integrating out $\tilde{K}$ above, one finds the distribution of $\omega_2$ for tails that satisfy $\omega_1=E_0$:
\begin{eqnarray}
\rho^{t+}_{\omega_2}(\omega_2;E,E_0) & = & \frac{1}{E_0^2}(E-\frac{E_0-\omega_2}{2}) \label{eq:rho_t_w2_pos}\\ 
 & & \quad\quad\quad E_0-2E < \omega_2 < E_0 \;. \nonumber
\end{eqnarray}
The corresponding distribution for tails that satisfy \mbox{$\omega_1=-E_0$} is 
\begin{eqnarray}
\rho^{t-}_{\omega_2}(\omega_2;E,E_0) & = & \frac{1}{E_0^2}(E-\frac{E_0+\omega_2}{2}) \label{eq:rho_t_w2_neg}\\
 & & \quad\quad\quad -E_0 < \omega_2 < -E_0+2E \;. \nonumber
\end{eqnarray}
The distributions in Eqs.~\eqref{eq:rho_t_w2_pos} and \eqref{eq:rho_t_w2_neg} are normalized to $(E/E_0)^2$, so that the survival probability is one when $E=E_0$, but decreases as $E$ decreases due to strong coupling decimation. Note that this probability can be deduced geometrically from Fig.~\ref{fig:rho_tail_flow}.

We can also write down the distribution of $\omega_1$ for tails that were made at $E_0$:
\begin{eqnarray}
\rho^{t+}_{\omega_1}(\omega_1;E,E_0) & = & \left(\frac{E}{E_0}\right)^2 \,\delta(\omega_1-E_0) \label{eq:rho_t_w1_pos}
\end{eqnarray}
\begin{eqnarray}
\rho^{t-}_{\omega_1}(\omega_1;E,E_0) & = & \left(\frac{E}{E_0}\right)^2 \,\delta(\omega_1+E_0) \label{eq:rho_t_w1_neg} \;,
\end{eqnarray}
for positive and negative $\omega_1$, respectively.

Now we find $\rho^t_{\omega}(\omega;E)$, the distribution of all tail frequencies at a given energy $E$. We integrate over all the possible energies $E_0$, at which tails could have been created. As the creation energy is decremented from $E_0$ to $E_0-dE_0$, $dE_0\cdot\rho^c_{\omega}(E_0;E_0)$ center pairs with positive $\omega_1$ are checked for fast oscillator decimation. An equal number of center pairs with negative $\omega_1$ are checked. Note that $\rho^c_{\omega}$ is given by Eq.~\eqref{eq:rho_c_w}. This leads to
\begin{widetext}
\begin{eqnarray}
\rho^t_{\omega}(\omega;E) & = & \frac{1}{2}\int_E^{\tilde{E}} dE_0 \,\rho^c_{\omega}(E_0;E_0)
\left[\,\rho^{t+}_{\omega_1}(\omega;E,E_0) + \rho^{t-}_{\omega_1}(\omega;E,E_0) + \rho^{t+}_{\omega_2}(\omega;E,E_0) + \rho^{t-}_{\omega_2}(\omega;E,E_0) \,\right] \\
 & \equiv & \tilde{\rho}^{t+}_{\omega_1}(\omega;E) + \tilde{\rho}^{t-}_{\omega_1}(\omega;E) + \tilde{\rho}^{t+}_{\omega_2}(\omega;E) + \tilde{\rho}^{t-}_{\omega_2}(\omega;E) \;.
\end{eqnarray}
\end{widetext}
The factor of $1/2$ accounts for the fact that half of the pairs checked for fast oscillator decimation will survive and become tails.

Then we calculate each of the terms:
\begin{eqnarray}
\tilde{\rho}^{t+}_{\omega_1}(\omega;E) & = & \frac{1}{2}\int_E^{\tilde{E}} dE_0 \, 
\frac{E_0^2}{\tilde{E}^3} \frac{E^2}{E_0^2} \,\delta(\omega_1-E_0) \nonumber \\
 & = & \frac{E^2}{2\tilde{E}^3} \quad\quad\quad\quad E<\omega<\tilde{E} \\
\tilde{\rho}^{t-}_{\omega_1}(\omega;E) & = & \frac{E^2}{2\tilde{E}^3} \quad\quad\quad -\tilde{E}<\omega<-E 
\end{eqnarray}
\begin{eqnarray}
\tilde{\rho}^{t+}_{\omega_2}(\omega;E) & = & \frac{1}{2}\int_{\mbox{\scriptsize{max}}(E,\omega)}^{\mbox{\scriptsize{min}}(\tilde{E},\omega+2E)} dE_0 \frac{E_0^2}{\tilde{E}^3} \frac{1}{E_0^2}\left(E-\frac{E_0-\omega}{2}\right) \nonumber \\
 & = & \left\{\begin{array}{cc}\frac{(\omega+E)^2}{8\tilde{E}^3} & \quad\quad\quad |\omega|<E \\ \frac{E^2}{2\tilde{E}^3} & \quad\quad\quad \omega>E \end{array}\right.
\end{eqnarray}
\begin{eqnarray}
 \tilde{\rho}^{t-}_{\omega_2}(\omega;E) & = & \frac{1}{2}\int_{\mbox{\scriptsize{max}}(E,-\omega)}^{\mbox{\scriptsize{min}}(\tilde{E},-\omega+2E)} dE_0 \frac{E_0^2}{\tilde{E}^3} \frac{1}{E_0^2}\left(E-\frac{E_0+\omega}{2}\right) \nonumber \\
 & = & \left\{\begin{array}{cc}\frac{(\omega-E)^2}{8\tilde{E}^3} & \quad\quad\quad |\omega|<E \\ \frac{E^2}{2\tilde{E}^3} & \quad\quad\quad \omega<-E \end{array}\right.
\end{eqnarray}
We have assumed that $\omega,E \ll \tilde{E}$. In other words, we are looking at low frequencies at low energies, i.e.\ when the RG is close to the unsynchronized fixed point. 

Summing up the four terms, we find the frequency distribution of tails in this regime:
\begin{eqnarray}
\rho^t_{\omega}(\omega;E) & = & \left\{\begin{array}{cc}\frac{E^2}{\tilde{E}^3} & |\omega|>E \\ \frac{\omega^2+E^2}{4\tilde{E}^3} & |\omega|<E \end{array}\right. \label{eq:rho_t_w}\;.
\end{eqnarray}
By comparing Eqs.~\eqref{eq:rho_c_w} and \eqref{eq:rho_t_w}, one sees that there are relatively more and more tails than centers as $E$ decreases. By integrating $\rho^c_{\omega}(\omega;E)$ and $\rho^t_{\omega}(\omega;E)$ on the interval $-E<\omega<E$, one finds that there are $2(E/\tilde{E})^3$ and $(2/3)(E/\tilde{E})^3$ oscillators with $|\omega|<E$ belonging to tails and centers, respectively. In otherwords, there are $(E/\tilde{E})^3$ center pairs and $(2/3)(E/\tilde{E})^3$ tail pairs with at least one oscillator satisfying $|\omega|<E$. Thus, $3/5$ of all such pairs are centers.

By adding on the frequency distribution of centers given by Eq.~\eqref{eq:rho_c_w}, we find the frequency distribution of all unisolated oscillators
\begin{eqnarray}
\rho_{\omega}(\omega;E) & = & \left\{\begin{array}{cc}\frac{E^2}{\tilde{E}^3} & |\omega|>E \\ \frac{\omega^2+5E^2}{4\tilde{E}^3} & |\omega|<E \end{array}\right. \;.
\end{eqnarray}
This matches the flat-quadratic distribution found with the numerical RG in Fig.~\ref{fig:w_k2mu_flow_iso}.

\bibliography{pre2009}

\begin{thebibliography}{20}
\expandafter\ifx\csname natexlab\endcsname\relax\def\natexlab#1{#1}\fi
\expandafter\ifx\csname bibnamefont\endcsname\relax
  \def\bibnamefont#1{#1}\fi
\expandafter\ifx\csname bibfnamefont\endcsname\relax
  \def\bibfnamefont#1{#1}\fi
\expandafter\ifx\csname citenamefont\endcsname\relax
  \def\citenamefont#1{#1}\fi
\expandafter\ifx\csname url\endcsname\relax
  \def\url#1{\texttt{#1}}\fi
\expandafter\ifx\csname urlprefix\endcsname\relax\def\urlprefix{URL }\fi
\providecommand{\bibinfo}[2]{#2}
\providecommand{\eprint}[2][]{\url{#2}}

\bibitem[{\citenamefont{Kuramoto}(1984)}]{kuramoto84}
\bibinfo{author}{\bibfnamefont{Y.}~\bibnamefont{Kuramoto}},
  \emph{\bibinfo{title}{Chemical Oscillations, Waves, and Turbulence}}
  (\bibinfo{publisher}{Springer-Verlag}, \bibinfo{address}{Berlin},
  \bibinfo{year}{1984}).

\bibitem[{\citenamefont{Pikovsky et~al.}(2001)\citenamefont{Pikovsky,
  Rosenblum, and Kurths}}]{pikovsky01}
\bibinfo{author}{\bibfnamefont{A.~S.} \bibnamefont{Pikovsky}},
  \bibinfo{author}{\bibfnamefont{M.}~\bibnamefont{Rosenblum}},
  \bibnamefont{and} \bibinfo{author}{\bibfnamefont{J.}~\bibnamefont{Kurths}},
  \emph{\bibinfo{title}{Synchronization: A Universal Concept in Nonlinear
  Science}} (\bibinfo{publisher}{Cambridge University Press},
  \bibinfo{address}{New York}, \bibinfo{year}{2001}).

\bibitem[{\citenamefont{Acebr\'{o}n et~al.}(2005)\citenamefont{Acebr\'{o}n,
  Bonilla, Vicente, Ritort, and Spigler}}]{acebron05}
\bibinfo{author}{\bibfnamefont{J.~A.} \bibnamefont{Acebr\'{o}n}},
  \bibinfo{author}{\bibfnamefont{L.~L.} \bibnamefont{Bonilla}},
  \bibinfo{author}{\bibfnamefont{C.~J.~P.} \bibnamefont{Vicente}},
  \bibinfo{author}{\bibfnamefont{F.}~\bibnamefont{Ritort}}, \bibnamefont{and}
  \bibinfo{author}{\bibfnamefont{R.}~\bibnamefont{Spigler}},
  \bibinfo{journal}{Rev.\ Mod.\ Phys.} \textbf{\bibinfo{volume}{77}},
  \bibinfo{pages}{137} (\bibinfo{year}{2005}).

\bibitem[{\citenamefont{Wiesenfeld et~al.}(1996)\citenamefont{Wiesenfeld,
  Colet, and Strogatz}}]{wiesenfeld96}
\bibinfo{author}{\bibfnamefont{K.}~\bibnamefont{Wiesenfeld}},
  \bibinfo{author}{\bibfnamefont{P.}~\bibnamefont{Colet}}, \bibnamefont{and}
  \bibinfo{author}{\bibfnamefont{S.~H.} \bibnamefont{Strogatz}},
  \bibinfo{journal}{Phys.\ Rev.\ Lett.} \textbf{\bibinfo{volume}{76}},
  \bibinfo{pages}{404} (\bibinfo{year}{1996}).

\bibitem[{\citenamefont{Vladimirov et~al.}(2003)\citenamefont{Vladimirov,
  Kozyreff, and Mandel}}]{vladimirov03}
\bibinfo{author}{\bibfnamefont{A.~G.} \bibnamefont{Vladimirov}},
  \bibinfo{author}{\bibfnamefont{G.}~\bibnamefont{Kozyreff}}, \bibnamefont{and}
  \bibinfo{author}{\bibfnamefont{P.}~\bibnamefont{Mandel}},
  \bibinfo{journal}{Europhys.\ Lett.} \textbf{\bibinfo{volume}{61}},
  \bibinfo{pages}{613} (\bibinfo{year}{2003}).

\bibitem[{\citenamefont{Varela et~al.}(2001)\citenamefont{Varela, Lachaux,
  Rodriguez, and Martinerie}}]{varela01}
\bibinfo{author}{\bibfnamefont{F.}~\bibnamefont{Varela}},
  \bibinfo{author}{\bibfnamefont{J.-P.} \bibnamefont{Lachaux}},
  \bibinfo{author}{\bibfnamefont{E.}~\bibnamefont{Rodriguez}},
  \bibnamefont{and}
  \bibinfo{author}{\bibfnamefont{J.}~\bibnamefont{Martinerie}},
  \bibinfo{journal}{Nature Reviews Neurosci.} \textbf{\bibinfo{volume}{2}},
  \bibinfo{pages}{229} (\bibinfo{year}{2001}).

\bibitem[{\citenamefont{N\'{e}da et~al.}(2000)\citenamefont{N\'{e}da, Ravasz,
  Brechet, Vicsek, and Barab\'{a}si}}]{neda00}
\bibinfo{author}{\bibfnamefont{Z.}~\bibnamefont{N\'{e}da}},
  \bibinfo{author}{\bibfnamefont{E.}~\bibnamefont{Ravasz}},
  \bibinfo{author}{\bibfnamefont{Y.}~\bibnamefont{Brechet}},
  \bibinfo{author}{\bibfnamefont{T.}~\bibnamefont{Vicsek}}, \bibnamefont{and}
  \bibinfo{author}{\bibfnamefont{A.-L.} \bibnamefont{Barab\'{a}si}},
  \bibinfo{journal}{Nature} \textbf{\bibinfo{volume}{403}},
  \bibinfo{pages}{849} (\bibinfo{year}{2000}).

\bibitem[{\citenamefont{Hong et~al.}(2005)\citenamefont{Hong, Park., and
  Choi}}]{hong05}
\bibinfo{author}{\bibfnamefont{H.}~\bibnamefont{Hong}},
  \bibinfo{author}{\bibfnamefont{H.}~\bibnamefont{Park.}}, \bibnamefont{and}
  \bibinfo{author}{\bibfnamefont{M.~Y.} \bibnamefont{Choi}},
  \bibinfo{journal}{Phys.\ Rev.\ E} \textbf{\bibinfo{volume}{72}},
  \bibinfo{pages}{036217} (\bibinfo{year}{2005}).

\bibitem[{\citenamefont{Hong et~al.}(2007)\citenamefont{Hong, Chat\'{e}, Park.,
  and Tang}}]{hong07}
\bibinfo{author}{\bibfnamefont{H.}~\bibnamefont{Hong}},
  \bibinfo{author}{\bibfnamefont{H.}~\bibnamefont{Chat\'{e}}},
  \bibinfo{author}{\bibfnamefont{H.}~\bibnamefont{Park.}}, \bibnamefont{and}
  \bibinfo{author}{\bibfnamefont{L.-H.} \bibnamefont{Tang}},
  \bibinfo{journal}{Phys.\ Rev.\ Lett.} \textbf{\bibinfo{volume}{99}},
  \bibinfo{pages}{184101} (\bibinfo{year}{2007}).

\bibitem[{\citenamefont{Goldenfeld}(1992)}]{goldenfeld92}
\bibinfo{author}{\bibfnamefont{N.}~\bibnamefont{Goldenfeld}},
  \emph{\bibinfo{title}{Lectures on Phase Transitions and the Renormalization
  Group}} (\bibinfo{publisher}{Westview Press}, \bibinfo{address}{Boulder, CO},
  \bibinfo{year}{1992}).

\bibitem[{\citenamefont{Sakaguchi et~al.}(1987)\citenamefont{Sakaguchi,
  Shinomoto, and Kuramoto}}]{sakaguchi87}
\bibinfo{author}{\bibfnamefont{H.}~\bibnamefont{Sakaguchi}},
  \bibinfo{author}{\bibfnamefont{S.}~\bibnamefont{Shinomoto}},
  \bibnamefont{and} \bibinfo{author}{\bibfnamefont{Y.}~\bibnamefont{Kuramoto}},
  \bibinfo{journal}{Prog.\ Theor.\ Phys.} \textbf{\bibinfo{volume}{77}},
  \bibinfo{pages}{1005} (\bibinfo{year}{1987}).

\bibitem[{\citenamefont{Daido}(1988)}]{daido88}
\bibinfo{author}{\bibfnamefont{H.}~\bibnamefont{Daido}},
  \bibinfo{journal}{Phys.\ Rev.\ Lett.} \textbf{\bibinfo{volume}{61}},
  \bibinfo{pages}{231} (\bibinfo{year}{1988}).

\bibitem[{\citenamefont{Strogatz and Mirollo}(1988)}]{strogatz88}
\bibinfo{author}{\bibfnamefont{S.~H.} \bibnamefont{Strogatz}} \bibnamefont{and}
  \bibinfo{author}{\bibfnamefont{R.~E.} \bibnamefont{Mirollo}},
  \bibinfo{journal}{J.\ Phys.\ A} \textbf{\bibinfo{volume}{21}},
  \bibinfo{pages}{L699} (\bibinfo{year}{1988}).

\bibitem[{\citenamefont{Kogan et~al.}()\citenamefont{Kogan, Rogers, Cross, and
  Refael}}]{kogan09}
\bibinfo{author}{\bibfnamefont{O.}~\bibnamefont{Kogan}},
  \bibinfo{author}{\bibfnamefont{J.~L.} \bibnamefont{Rogers}},
  \bibinfo{author}{\bibfnamefont{M.~C.} \bibnamefont{Cross}}, \bibnamefont{and}
  \bibinfo{author}{\bibfnamefont{G.}~\bibnamefont{Refael}},
  \bibinfo{note}{accepted for publication in Phys.\ Rev.\ E, arXiv:0810.3075}.

\bibitem[{\citenamefont{Kogan}(2008)}]{kogan08}
\bibinfo{author}{\bibfnamefont{O.}~\bibnamefont{Kogan}}, Ph.D. thesis,
  \bibinfo{school}{California Institute of Technology} (\bibinfo{year}{2008}).

\bibitem[{\citenamefont{Press et~al.}(1992)\citenamefont{Press, Flannery,
  Teukolsky, and Vetterling}}]{press92}
\bibinfo{author}{\bibfnamefont{W.~H.} \bibnamefont{Press}},
  \bibinfo{author}{\bibfnamefont{B.~P.} \bibnamefont{Flannery}},
  \bibinfo{author}{\bibfnamefont{S.~A.} \bibnamefont{Teukolsky}},
  \bibnamefont{and} \bibinfo{author}{\bibfnamefont{W.~T.}
  \bibnamefont{Vetterling}}, \emph{\bibinfo{title}{Numerical Recipes in C: The
  Art of Scientific Computing}} (\bibinfo{publisher}{Cambridge University
  Press}, \bibinfo{address}{Cambridge}, \bibinfo{year}{1992}).

\bibitem[{\citenamefont{Fisher}(1994)}]{fisher94}
\bibinfo{author}{\bibfnamefont{D.}~\bibnamefont{Fisher}},
  \bibinfo{journal}{Phys.\ Rev.\ B} \textbf{\bibinfo{volume}{50}},
  \bibinfo{pages}{3799} (\bibinfo{year}{1994}).

\bibitem[{\citenamefont{Sondhi et~al.}(1997)\citenamefont{Sondhi, Girvin,
  Carini, and Shahar}}]{sondhi97}
\bibinfo{author}{\bibfnamefont{S.~L.} \bibnamefont{Sondhi}},
  \bibinfo{author}{\bibfnamefont{S.~M.} \bibnamefont{Girvin}},
  \bibinfo{author}{\bibfnamefont{J.~P.} \bibnamefont{Carini}},
  \bibnamefont{and} \bibinfo{author}{\bibfnamefont{D.}~\bibnamefont{Shahar}},
  \bibinfo{journal}{Rev.\ Mod.\ Phys.} \textbf{\bibinfo{volume}{69}},
  \bibinfo{pages}{1} (\bibinfo{year}{1997}).

\bibitem[{\citenamefont{Ochab and G\'{o}ra}()}]{ochab09}
\bibinfo{author}{\bibfnamefont{J.}~\bibnamefont{Ochab}} \bibnamefont{and}
  \bibinfo{author}{\bibfnamefont{P.~F.} \bibnamefont{G\'{o}ra}},
  \bibinfo{note}{submitted to Acta Phys.\ Pol.\ B, arXiv:0909.0043}.

\bibitem[{pro()}]{proof1}
\bibinfo{note}{By the definition of a center, $|\Delta\omega|<2E$. When a
  center gets checked for strong coupling decimation, $\tilde{K}=E$. Thus,
  $r=2\tilde{K}/|\Delta\omega|>1$. Hence, when a center is checked for strong
  coupling decimation, it will always be decimated as such.}

\end{thebibliography}

\end{document}